\documentstyle[prb,aps]{revtex}
\input epsfig.sty

\begin{document}
\draft
\title{
\flushright {\tt to appear in Phys. Rev. B } \protect \\
Phase transitions in a frustrated XY model with zig-zag couplings}
\author{Mourad Benakli $^{\dagger }$ and Enzo Granato $^{\dagger \ddagger }$}
\address{$^{\dagger }$ Condensed Matter Physics Group, \\
International Centre for Theoretical Physics, \\
34100 Trieste, Italy \\
$^{\ddagger }$ Laborat\'orio Associado de Sensores e Materiais,\\
Instituto Nacional de Pesquisas Espaciais, \\
12225 - S\~ao Jos\'e dos Campos, SP, Brasil }
\maketitle

\begin{abstract}
We study a new generalized version of the square-lattice frustrated XY model
where unequal ferromagnetic and antiferromagnetic couplings are arranged in
a zig-zag pattern. The ratio between the couplings $\rho$ can be used to
tune the system, continuously, from the isotropic square-lattice to the
triangular-lattice frustrated XY model. The model can be physically realized
as a Josephson-junction array with two different couplings, in a magnetic
field corresponding to half-flux quanta per plaquette. Mean-field
approximation, Ginzburg-Landau expansion and finite-size scaling of Monte
Carlo simulations are used to study the phase diagram and critical behavior.
Depending on the value of $\rho$, two separate transitions or a transition
line in the universality class of the XY-Ising model, with combined $Z_2$
and $U(1)$ symmetries, takes place. In particular, the phase transitions of
the standard square-lattice and triangular-lattice frustrated XY models
correspond to two different cuts through the same transition line. Estimates
of the chiral ($Z_2$) critical exponents on this transition line deviate
significantly from the pure Ising values, consistent with that along the
critical line of the XY-Ising model. This suggests that a frustrated XY
model or Josephson-junction array with a zig-zag coupling modulation can
provide a physical realization of the XY-Ising model critical line.
\end{abstract}

\pacs{75.40 Mg, 64.60 Cn, 74.40+k}



\section{Introduction}

There has been an increasing interest in frustrated XY models in relation to
Josephson-junction arrays in a magnetic field \cite{tj83,halsey85,gkn96}. At
a particular value of the external field, corresponding to half flux quanta
per plaquette of the array, the ideal system is isomorphic to a frustrated
XY model, or Villain's odd model \cite{villain}, with ferromagnetic and
antiferromagnetic bonds satisfying the odd rule, in which every plaquette
has an odd number of antiferromagnetic bonds. Frustration has the effect of
introducing a discrete $Z_2$ symmetry in the ground state with an associated
chiral (Ising-like) order parameter, in addition to the continuous $U(1)$
symmetry. The interplay between these two order parameters may lead to
critical behavior which is not present in the unfrustrated model which is
known to have a transition in the Kosterliz-Thouless (KT) universality class.

Earlier Monte Carlo simulation results for the isotropic square-lattice
(SFXY) \cite{tj83,bergediep} and triangular-lattice (TFXY) \cite{ms84,ljnl86}
frustrated XY model, and some recent ones \cite{grest,olsson}, suggest a
critical behavior associated with the chiral order parameter in agreement
with pure Ising exponents while the continuous (XY) degrees of freedom
display the main features of the KT transition, possibly with a nonuniversal
jump. Estimates of the corresponding critical temperatures are always too
close to be satisfactorily resolved within the errorbars, specially when
possible systematic errors due to the assumed KT scaling forms are taken
into account. These results can either be regarded as an indication of a
single but decoupled transition, where the Ising and XY variables have
standard behavior and the same critical point, or else there are two
separate by close transitions of Ising and KT type. There exist also some
appealing arguments which exclude one of the two possibilities, Ising
followed by a KT transition for increasing temperature, in the case of a
double transition scenario \cite{halsey85}. Other numerical works, however,
which attempt an improved estimate of the chiral critical exponents tend to
conclude that these exponents deviate significantly from the pure Ising
values \cite{lkg91,jose92,knops94}. In particular, based on the results for
the coupled XY-Ising model as an effective Hamiltonian for these systems 
\cite{eg87,gkln91}, it has been argued that, in the case of the single
transition scenario, both the SFXY and TFXY model display a transition with
exponents deviating from the pure Ising values. Moreover, the exponents are
given by the corresponding values along the critical line of this model.
Estimates of chiral exponents from Monte Carlo data \cite{lgk91} and
transfer-matrix calculations \cite{gn93,knops94} are consistent with the
XY-Ising model universality class \cite{lgk91,ngk95}.

A generalized version of the SFXY model has been introduced by Berge {et al.}
\cite{bergediep} where the strength of the antiferromagnetic bonds can be
varied. This introduces a particular anisotropy into the system and leads to
clearly separated Ising and KT-like transitions for unequal strengths \cite
{gk86,Gabay89,eikmans} but which appear to merge into a single one for equal
strengths, corresponding to the isotropic SFXY model. There is a critical
value for the bond strength, $1/3$, below which the twofold degeneracy
disappears. Other generalizations have been introduced for the TFXY model
that also leads to a critical strength below which the frustration effect is
suppressed \cite{zsgb93}. A common feature in the topology of the phase
diagram of these generalized versions is that the isotropic model always
corresponds to the region where chiral and XY ordering can not be clearly
resolved. However, so far, the SFXY and TFXY models have been treated as
separated models.

In this work, we introduce a new generalized version of the SFXY model where
unequal ferromagnetic and antiferromagnetic couplings are arranged in a
zig-zag pattern. The ratio between the couplings $\rho $ can be used to tune
the system, continuously, from the isotropic SFXY to the TFXY model,
allowing the study of both models within the same framework. The model can be
physically realized as a Josephson-junction array with two different
couplings, in a magnetic field corresponding to half-flux quanta per
plaquette. We use a mean-field approximation, Ginzburg-Landau expansion and
finite-size scaling of Monte Carlo simulations to study the phase diagram
and critical behavior. Depending on the value of $\rho $, two separate
transitions or a transition line with combined $Z_2$ and $U(1)$ symmetries,
takes place. Based on an effective Hamiltonian, we show that this transition
line is in the universality class of the XY-Ising model and the phase
transitions of the standard SFXY and TFXY models correspond to two different
cuts through the same transition line. Estimates of the chiral ($Z_2$)
critical exponents are consistent with that along the critical line of the
coupled XY-Ising model, suggesting a possible physical realization of the
XY-Ising model critical line in a frustrated XY model or Josephson-junction
array with a zig-zag coupling modulation.

The remainder of the paper is organized as follows. In Sec. II, we define
the model. In Sec. III, the ground state properties obtained by two
different methods are presented. In Sec. IV, a mean field approximation is
used to obtain the global features of the phase diagram. In Sec. V, the
effective Hamiltonian obtained by Guinzburg Landau expansions is presented
and its relation to coupled XY models and the XY-Ising model is discussed.
In Sec. VI, we present numerical results of Monte Carlo simulations for the
phase diagram and chiral critical exponents obtained from finite-size
scaling. Finally, Sec. VII is devoted to the conclusions.

\section{The model}

\begin{figure}[h]
\centering\epsfig{file=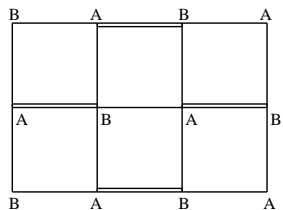,angle=-90,bbllx=1cm,bblly=1cm,bburx=20cm,
bbury=28cm,width=8cm}

\caption{Generalized frustrated XY model with zig-zag coupling modulation.
Continuous lines correspond to coupling $J_{ij}=J$ and double lines to $%
J_{ij}=J^{\prime }$. The sites A and B denote two sublattices where spins do
not interact. }
\label{zz-model.eps}
\end{figure}

The generalized version of the frustrated XY model introduced by Berge {et
al.} \cite{bergediep} can be regarded as an XY version of one of the two
frustrated Ising models with periodic interactions first introduced by
Andr\'e {et al.} \cite{andre}. The other model has the important feature
that it reduces to the triangular-lattice antiferromagnetic Ising model in
one particular limit. In analogy to this model, we consider a system of
classical XY spins on a square lattice with nearest neighbors interactions
modulated in a periodic pattern. The Hamiltonian of this zig-zag model is
given by

\begin{equation}
{\cal H}=-\frac 12\sum_{<i,j>}J_{ij}\vec S_i \cdot \vec S_j ,
\label{Hamil-ss}
\end{equation}
where the sum is restricted to the first neighbors and $\vec S_i$ is a
two-component unit vector. The couplings $J_{ij}$ can have two different
values, $J$ and $J^\prime $, distributed periodically in a zig-zag pattern
as indicated in Fig. \ref{zz-model.eps}. We choose $J$ to be ferromagnetic ($%
J>0$) and define $J^\prime = - \rho J$, where $\rho$ is the coupling ratio.
We are interested in the case $\rho > 0$, where each plaquette has an odd
number of antiferromagnet bonds, Villain's odd rule \cite{villain}, which
leads to frustration effects.

When $\rho =1$, the model reduces to the SFXY model while in the limit $\rho
\rightarrow +\infty $ it is topologically equivalent to the TFXY model. The
latter limit can be easily established after performing the "gauge"
transformation $S_i \rightarrow \epsilon_i S_i$, $J_{i,j} \rightarrow J_{ij}
\epsilon_i \epsilon_j$, where $\epsilon_i = 1$ and $-1$ on the sublattices A
and B of Fig. \ref{zz-model.eps}, respectively, resulting in
antiferromagnetic $J$ and ferromagnetic $J^\prime$ couplings. When $\rho
\rightarrow +\infty$, each pair of spins connected by a $J^\prime$ bond
become locked and may be replaced by an effective spin, leading to an
antiferromagnetic XY model with the same coordination number as the
triangular lattice. The model is then well suited for the study of the
universality classes of both SFXY and TFXY models. When $\rho =0$, Eq. (\ref
{Hamil-ss}) reduces to a ferromagnetic XY model on a hexagonal lattice which
undergoes a KT transition.

\section{Ground state}

\begin{figure}[h]
\centering\epsfig{file=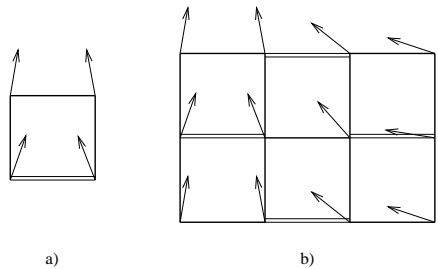,angle=-90,bbllx=1cm,bblly=1cm,bburx=20cm,
bbury=28cm,width=8cm}

\caption{Ground state for $\rho >1/3$ consisting in a helical configuration
of spins. a) ground state configuration of a single plaquette and b) spin
configuration for the infinite lattice. }
\label{min-h.eps}
\end{figure}

In another generalization of the frustrated XY model \cite{bergediep}, it
has been shown that the lowest energy state can be constructed by building
up the configuration of the infinite lattice from the ground state
configuration of a single plaquette. In our case, the same procedure can be
used if we allow for rotations and reflections of the ''one plaquette ground
state configuration'' which also assures that the true ground state is
obtained. No assumption on the periodicity of the ground state is made. The
plaquette configuration is indicated in Fig. \ref{min-h.eps}a for $\rho >1/3$
and it is the same as used in Ref. \onlinecite{bergediep}. The spin
configuration is collinear for $\rho <1/3$ and a canted one for $\rho >1/3$.
For the canted configuration one can define a chiral variable for each
plaquette $P$ 
\begin{equation}
\sigma _P=\frac 1{\sigma _o}\sum_{<ij>\in P}J_{ij}\vec S_i\times \vec S_j,
\label{chi}
\end{equation}
where $\sum_{<ij>\in P}$ is a direct summation around the plaquette and $%
\sigma _o$ is a normalization factor given by

\begin{equation}
\sigma _o=\frac{3\rho +1}{2\rho }\sqrt{\frac{3\rho -1}\rho }
\end{equation}

For $\rho >1/3$, the ground state of the infinite system constructed by the
above procedure consists in a helical spin ordering which is {\it %
incommensurate} with the underlying square lattice, except when $\rho = 1$
and $\rho = \infty $, corresponding to the SFXY and TFXY models. The pitch $%
\Delta$ of the helical configuration, can be obtained from half the phase
difference within the same sublattice in the $\hat x$ direction and is given
by

\begin{equation}
\Delta =2\cos ^{-1}(\frac 12\sqrt{\frac{\rho +1}\rho })=\arccos \left( \frac{%
1-\rho }{2\rho }\right)
\end{equation}

In Fig. \ref{min-h.eps}b we show the resulting ground state configuration.
The ground state is double degenerate, corresponding to an antiferromagnetic
arrangement of plaquette chiralities $\sigma = \pm 1 $.

For $\rho <1/3$, where the single plaquette configuration is collinear, the
ground state is a ferromagnetic configuration of spins.

As an alternative to the above method, the ground state can also be obtained
by a direct minimization of the Fourier-transform interaction matrix, $%
-J_{q,q^{\prime }}$. In the present case, we note that there are two
non-interacting sublattices, corresponding to the sites A and B in Fig. \ref
{zz-model.eps}, where the Fourier transform can be easily carried out. The
interaction matrix $J_q^{k.l}$, where $k,l$ denote the sublattices A and B,
can be written as 
\begin{equation}
J_q^{k.l}=J\left( 
\begin{array}{cc}
0 & e^{iq_x}-\rho e^{-iq_x}+e^{iq_y}+e^{-iq_y} \\ 
-\rho e^{iq_x}+e^{-iq_x}+e^{iq_y}+e^{-iq_y} & 0
\end{array}
\right)  \label{Jqq}
\end{equation}
The eigenvalues are given by, $\lambda _q=\pm V_q$, where 
\begin{equation}
V_q=J\sqrt{\left( 1+\rho \right) ^2+4(-\rho \cos ^2(q_x)+\cos
^2(q_y)+(1-\rho )\cos (q_x)\cos (q_y))},  \label{eigen}
\end{equation}
and the dominant eigenvalue $\lambda _q=+V_q$ reaches a maximum at 
\begin{equation}
\begin{tabular}{ll}
$\left( q_x=0,q_y=0\right) $ & for $\rho \leq \frac 13$ \\ 
$\left( q_x=\pm \arccos \left[ \left( 1-\rho \right) /\left( 2\rho \right)
\right] ,q_y=0\right) $ & for $\rho \geq \frac 13$%
\end{tabular}
\label{minv}
\end{equation}
From Eq. (\ref{minv}), the wavevector characterizing the ground state for $%
\rho >1/3$ is, in general, incommensurate with the lattice periodicity in
the $\hat x$ direction, except for $\rho =1$ and $\rho =\infty $
corresponding to the SFXY and TFXY models. The eigenvector associated to the
largest eigenvalue is a possible realization of ground state provided the
corresponding spin configuration satisfy the unit vector condition $|\vec S%
_i|=1$. In the present case, they do satisfy this condition and correspond
to the same configuration as found from the single plaquette method
described above.

\section{Mean-field phase diagram}

The general form of the phase diagram can be obtained by a mean field (MF)
analysis. Although, at finite temperatures, this analysis neglects the role
of fluctuations it gives nevertheless a good qualitative picture of the
phase diagram that can also be greatly improved by perturbative or
variational techniques. The details are described in the Appendix.

\begin{figure}[h]
\centering\epsfig{file=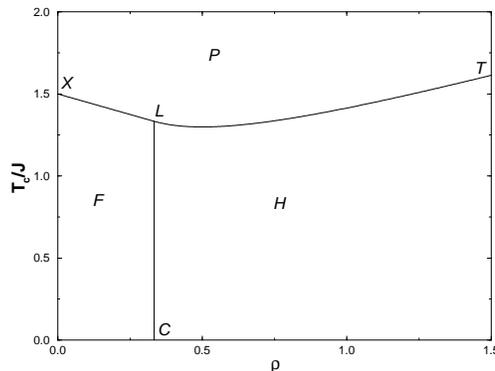,angle=-90,bbllx=1cm,bblly=1cm,bburx=20cm,
bbury=28cm,width=8cm}

\caption{Mean field phase diagram. F indicates the ferromagnetic phase, P
the paramagnetic phase and H the helical (chiral) phase.}
\label{tc-mf.eps}
\end{figure}

Fig. \ref{tc-mf.eps} shows the phase diagram obtained by the mean field
approximation. For $\rho <1/3$, the system undergoes a transition from a
paramagnetic to a ferromagnetic phase along the transition line XL. This
transition is in the KT universality class since there is only a single
critical mode $q=(0,0)$ and no additional symmetry in the ground state
besides the continuous $U(1)$ symmetry. For $\rho >1/3$, there is a
paramagnetic phase at high temperatures and a helical phase at low
temperatures which is incommensurate with the lattice periodicity except for 
$\rho =1$ and $\rho +\infty $ where the model reduces to the SFXY and TFXY
models, respectively. The helical phase has an additional discrete $Z_2$
symmetry associated with the antiferromagnetic arrangement of plaquette
chiralities $\sigma _P$ in the ground state. The mean field analysis gives a
single transition for $\rho >1/3$. Therefore, the whole line LT for $\rho >1$%
, including the TFXY limit, is expected to have the same type of behavior as
the SFXY model at $\rho =1$. The nature of this transition however can not
be studied at mean field level and other methods are required, as will be
presented in Sec. V and VI. Note that, in contrast to the generalized
version of the frustrated XY model considered by Berge {\it et al.} \cite
{bergediep}, where a clear separation into two transitions for $\rho \ne 1$
is already found at the mean field level \cite{Gabay89}, the zig-zag model
displays two transitions only for $\rho <1/3$ within the same kind of
approximation. This suggests that the separation of these transitions is not
simply a result of the induced anisotropy for $\rho \ne 1$ but should be
related to the nature of the coupling between chiral (Ising) and XY degrees
of freedom, in agreement with arguments based on an effective coupled
XY-Ising model Hamiltonian \cite{gkln91}. As will be shown in Sec. IV, for
the zig-zag model the form of this coupling is unchanged for $\rho \sim 1$
and $\rho >1$, suggesting that a clear separation is not expected.

The transition line CL separating the ferromagnetic from helical phase can
be regarded as a commensurate-incommensurate transition which joins the
other transition lines, XL and LT, at a Lifshitz point $L$ at $T \ne 0$. In
mean field, this transition line is given by $\rho =1/3$ corresponding to
the stability boundary between the two modes in Eq. (\ref{minv}). Although,
there are interesting questions regarding the precise location of the
Lifshitz point and the nature of the phase transition along this line \cite
{gd80,SaslowCLF,benak1,benak3}, these will not be the subject of a detailed
study in this work.

\section{Effective Hamiltonian}

The universality class of phase transitions can be considered on the basis
of an effective Hamiltonian obtained by Ginzburg-Landau expansions. Invoking
the universality hypothesis, one expects that models with the same effective
Hamiltonian differing only by irrelevant terms are in the same universality
class. In this section, we discuss the critical behavior in the region $\rho
>1/3$ where the ground state is double degenerated by deriving the
corresponding effective Hamiltonian. An effective Hamiltonian can be
obtained from the free energy functional, describing fluctuations around the
MF solution discussed in Sec. IV, via a Hubbard-Stratonovich transformation
in a standard way \cite{cd85,kawa88}. One replaces Eq. (\ref{Hamil-ss}) by 
\begin{equation}
\frac{{\cal F}}{kT}=\frac 12\sum_{i,j}K_{ij}^{-1}\vec t_i\cdot \vec t%
_j-\sum_iW(|\vec t_i|),  \label{Hamil-tt}
\end{equation}
where $\vec t_i$ are unconstrained spins weighted by $W(x)\sim
x^2/4-x^4/64+O(x^6)$ and $K_{iij}=J_{ij}/kT$. In the present case, we can
separate the lattice spins into two non-interacting sublattices,
corresponding to sites A and B in Fig. \ref{zz-model.eps}. The interaction
matrix $J_q^{l,k}$ is then given by Eq. (\ref{Jqq}) and the corresponding
eigenvalues by Eq. (\ref{eigen}). For $\rho >1/3$, there are two degenerated
modes $\vec \phi _{Q^{-}}$ and $\vec \phi _{Q^{+}}$, that maximizes the
dominant eigenvalue $\lambda =+V_Q$, corresponding to the wave vectors in
Eq. (\ref{minv}). Retaining these modes only and introducing the real
two-component fields $\vec \phi _1=\frac 12(\vec \phi _{Q^{-}}+\vec \phi
_{Q^{+}})$ and $\vec \phi _2=\frac 12i(\vec \phi _{Q^{-}}-\vec \phi
_{Q^{+}}) $ , one can expand Eq. (\ref{Hamil-tt}) to quartic order in $\vec 
\phi_{1,2}$ leading, in the continuum limit, to a free energy density of the
form 
\begin{eqnarray}
\beta f=\frac 12r_o(\vec \phi _1^2+\vec \phi _2^2)+\frac 12e[(\frac \partial
{\partial x}\vec \phi _1)^2+(\frac \partial {\partial x}\vec \phi _2)^2)]+%
\frac 12f[(\frac \partial {\partial y}\vec \phi _1)^2+(\frac \partial {%
\partial y}\vec \phi _2)^2]+\cr u(\vec \phi _1^2+\vec \phi _2^2)^2+v((\vec 
\phi _1\cdot \vec \phi _2)^2-\vec \phi _1^2\vec \phi _2^2),  \label{GL}
\end{eqnarray}
where $r_0=kT/\lambda _Q-1/2$, $e=\frac{kT}{2\lambda _Q^2}\frac{\partial ^2}{%
\partial q_x^2}\lambda _Q$ , $f=\frac{kT}{2\lambda _Q^2}\frac{\partial ^2}{%
\partial q_y^2}\lambda _Q$ , and $u,v>0$. For $\rho <1/3$, there is only one
critical mode, $(q_x,q_y)=(0,0)$, and the resulting Ginzburg-Landau
expansion has a single two-component fluctuating field which is known to lie
in the KT universality class. Apart from the space anisotropy, $e\ne f$ when 
$\rho \ne 1$, that can be eliminated by rescaling the $x$ and $y$ space
directions appropriately, the free energy (\ref{GL}) has the same form as
those obtained for the SFXY and TFXY \cite{cd85,yd85,kawa88,lgk91} in terms
of complex scalar fields $\varphi _i=|\varphi |e^{i\theta _i}$ . In
particular, since the present model incorporates both the SFXY and TFXY as
special cases, it clearly demonstrates that the SFXY and TFXY are described
by the same Ginzburg-Landau free energy up to quartic order, in agreement
with the arguments of Ref. \onlinecite{lgk91}.

As usual, in two dimensions, fluctuations in the magnitude of the order
parameter are assumed to be irrelevant. We can then approximate these
magnitudes by their corresponding mean-field values $|\psi _{1,2}|=\psi
_o=-r_o(2u-v)$ and consider only fluctuations of the phase $\theta _i$ in
Eq. (\ref{GL}), leading to an effective lattice Hamiltonian in the form of
two coupled XY models 
\begin{equation}
\beta {\cal H}=-\sum_{<ij>}[\Gamma _1\cos (\theta _{1,i}-\theta
_{1,j})+\Gamma _2\cos (\theta _{2,i}-\theta _{2,j})]-h\sum_i\cos 2(\theta
_{1,i}-\theta _{2,i})  \label{cxy}
\end{equation}
where $\Gamma _1=\Gamma _2=|\psi _o|\sqrt{ef}$ and the spatial anisotropy
has been removed by rescaling $x\rightarrow x\sqrt{e/f}$, $y\rightarrow y$.
In a renormalization study of this model \cite{gk86}, the $\Gamma _1=\Gamma
_2$ subspace is only preserved under renormalization if they are initially
equal. For $\Gamma _1\ne \Gamma _2$, a double transition is found with an
Ising followed by a KT transition as temperature is increased. It is also
found that the coupling term $h$ is a relevant variable locking the phase
difference into $\theta _{2i}=\theta _{1i}+\pi \tau $, where $\tau =0,1$.
This leads, in the $h\rightarrow \infty $, to an effective Hamiltonian in
the form of coupled XY and Ising models \cite{gkln91,lgk91} 
\begin{equation}
\beta H=-\sum_{<ij>}[(A_{eff}+B_{eff}\sigma _i\sigma _j)\cos (\theta
_i-\theta _j)+C_{eff}\sigma _i\sigma _j],  \label{cxyi}
\end{equation}
where $A_{eff}$, $B_{eff}$ and $C_{eff}$ are effective couplings which
depend on the initial values of $\Gamma _{1,2}$, $h$ and other couplings
generated by the renormalization procedure, and $\sigma _i=2\tau _i-1=\pm 1$
is an Ising-like variable. The condition $A_{eff}=B_{eff}$ is preserved if $%
\Gamma _1=\Gamma _2$ in Eq. (\ref{cxy}) as is the case for the zig-zag
model, even though this model is anisotropic for $\rho \ne 1$. This should
be contrasted to the generalized SFXY model considered by Berge {\it et al.} 
\cite{bergediep} where the Ginzburg-Landau free energy has the same form as
in Eq. (\ref{GL}) but with a spatial anisotropy, in the $x$ and $y$
directions, of different magnitudes for the $\vec \phi _1$ and $\vec \phi _2$
fields which can not be removed by simple rescaling \cite{gk86}. This leads
to coupled XY models with $\Gamma _1\ne \Gamma _2$ in Eq. (\ref{cxy}) and
consequently should be described by an XY-Ising model with $A\ne B$ in Eq. (%
\ref{cxyi}) which undergoes two separate transitions, an Ising followed by a
KT transition for increasing temperatures, in agreement with simulations 
\cite{bergediep}.

The phase diagram of the XY-Ising model of Eq. (\ref{cxyi}) for $A=B$
consists of three branches which meet at a multicritical point \cite{lgk91}.
One of the branches corresponds to single transitions with simultaneous loss
of XY and Ising order, and the other two to separate KT and Ising
transitions. The line of single transitions eventually becomes first order
further away from the branch point. Our model corresponds to a particular
path through the phase diagram of the XY-Ising model and the single or
double character of the transition depends on the relative position to the
multicritical point. Since there are already indications from numerical
simulations \cite{lkg91,jose92,knops94} that both SFXY and TFXY limits are
in the single transition region, we expect that the whole transition line LT
for $\rho >\rho _L$ in Fig. \ref{tc-mf.eps} should correspond to this
critical line. Numerical estimates of critical exponents associated with the 
$Z_2$ order parameter for the XY-Ising model deviate significantly from the
pure Ising values along the critical line \cite{lgk91,ngk95} and will be
used in Sec. VI to identify which particular path through the phase diagram
is realized for the zig-zag model.

\section{Monte Carlo simulations}

Due to the presence of an incommensurate phase, the standard {\it periodic
boundary conditions} are not appropriate for the zig-zag model since they
cause an additional frustration in the system. Therefore we use a {\it %
self-consistent boundary condition} that allow the system to adapt the
boundary condition to the pitch of the helical configuration \cite
{SaslowCLF,benak1}. In addition, this boundary condition also improves the
determination of the spin stiffness. A similar method \cite{olsson} has also
been used recently for the SFXY.

\subsection{Phase diagram}

To determine the global phase diagram we used simulations of a $36\times 36$
system for various values of $\rho $. For each value, two separate
simulations, one starting from the ground state and the other from the high
temperature phase, were used to estimate the critical temperature. The
transition temperature, $T_I$, associated with the chiral order parameter,
was obtained from the peak in the chiral staggered susceptibility, with the
chiral order parameter defined by Eq. (\ref{chi}). An estimate of the KT
transition temperature, $T_{KT}$, was obtained from the expected universal
value of the spin stiffness $\gamma $, $\gamma (T_{KT})/kT_{KT}=2/\pi $ , at
the transition. Since for $\rho \ne 1$ the model is anisotropic, $\gamma $
was obtained as $\gamma =\sqrt{\gamma _{xx}\gamma _{yy}}$, where $\gamma
_{xx}$ and $\gamma _{yy}$ are the $x$ and $y$ components of the stiffness $%
\gamma _{ij}$. This is the quantity that should be universal at the KT
transition. The anisotropy of the renormalized Gaussian model, at the
critical point, can be removed by rescaling the coordinate axes, as in the
derivation of Eq. (\ref{cxy}), leading to the geometric mean as the
effective stiffness. The same averaging procedure has also been used in Ref. %
\onlinecite{eikmans} to treat the anisotropic model of Ref. %
\onlinecite{bergediep}.

\begin{figure}[h]
\centering\epsfig{file=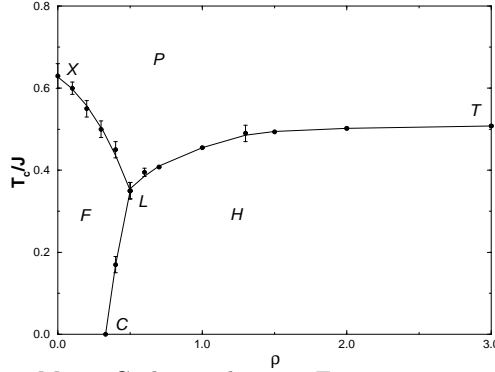,angle=-90,bbllx=1cm,bblly=1cm,bburx=20cm,
bbury=28cm,width=8cm}

\caption{Phase diagram obtained from Monte Carlo simulations. For $\rho >
\rho_l$, critical-temperature estimates $T_I$ and $T_{KT}$ agree within
errorbars and only $T_I$ is indicated. }
\label{graph.eps}
\end{figure}

The phase diagram obtained by Monte Carlo simulations is shown in Fig. \ref
{graph.eps}. The estimates of $T_{I}$ and $T_{KT}$ agree within the
errorbars for $\rho $ larger than a critical value $\rho_L$, which we take
as an estimate of the Lifshitz point. This phase diagram is similar to the
MF result of Sec. III but the Lifshitz point is located at $\left( \rho
_L\simeq 0.6,T_L\simeq 0.35\right) $ and should be compared with the MF
result, $\left( \rho =\frac 13,T=\frac 76\right) $. The transition line XL
has the main features of a KT transition with a jump in the spin stiffness
consistent with the universal value $2/\pi$ and a nondivergent specific heat.

\begin{figure}[h]
\centering\epsfig{file=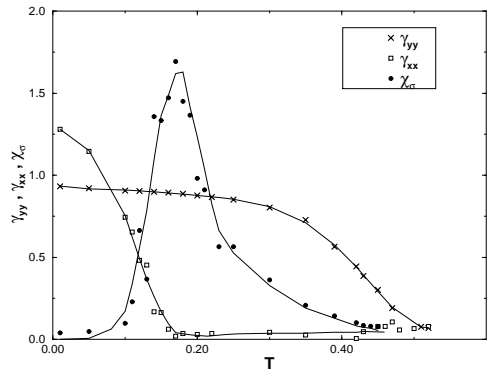,angle=-90,bbllx=1cm,bblly=1cm,bburx=20cm,
bbury=28cm,width=8cm}

\caption{Temperature dependence of spin stiffness and chiral susceptibility
through the CL transition line of the phase diagram in Fig. \ref{graph.eps}
at $\rho =0.4$. The data points for $\gamma_{xx}$ are scaled by $10$ and for 
$\chi_{\sigma}$ scaled by $1/10$.}
\label{gamma.eps}
\end{figure}

The transition line CL is characterized by a divergent chiral susceptibility
and an apparent continuous vanishing of $\gamma _{xx}$ while $\gamma _{yy}$
remains finite as shown in Fig. \ref{gamma.eps}. An analysis similar to the
one used in Ref. \onlinecite{eikmans} for the model studied by Berge {et al.}
\cite{bergediep} can also be applied to the zig-zag model and shows that $%
\gamma _{xx}$ is inversely proportional to the chiral susceptibility and
should therefore decrease continuously at the transition when the
susceptibility diverges. Similar behavior has been found in a generalized
model for the triangular lattice \cite{SaslowCLF,benak1}.

\subsection{Critical exponents}

There have been recently many attempts to obtain improved estimates of the
critical exponents for the fully frustrated XY model \cite
{tk90,lkg91,jose92,gn93,knops94}. For the continuous symmetry, the available
scaling forms requires the simultaneous fit of two or more parameters and an
assumption of KT behavior. This may lead to systematic errors in the
location of the KT transition temperature. For the chiral ( Ising-like)
order parameter there exist scaling analysis which do not require a precise
knowledge of the bulk $T_c$ and can provide an estimate of the critical
exponents with a one-parameter fit. As simple estimates of $T_I$ and $T_{KT}$
already agree within errorbars along the transition line for $\rho >\rho _L$%
, as indicated in Fig. \ref{graph.eps}, attempting to locate the transition
line using, separately, KT scaling forms for the $U(1)$ symmetry and pure
Ising critical behavior for the chiral variables will inevitably lead to
estimates of critical points which are difficult to resolve on purely
numerical grounds due to errorbars. However, if the critical behavior along
this line is in fact in the same universality class as the XY-Ising model as
suggested by the analysis of Sec. V, then in order to verify the single
nature of the transition, it is sufficient to study the $Z_2$ degrees of
freedom \cite{gkln91}. If the critical exponents are inconsistent with pure
Ising values, the transition cannot correspond to the Ising branch of a
double transition or to a single but decoupled transition. Moreover, the
value of the critical exponent can be used to verify if indeed the critical
behavior corresponds to the critical line of the XY-Ising model. In order to
estimate the chiral critical exponents independently of $T_c$, we use the
same method, based on the finite-size scaling of free energy barriers, which
has been applied to the SFXY and TFXY models \cite{lgk91}.

In order to obtain good statistics, we consider only systems of size $
8$ to $36\times 36$, with typically $6-12\times 10^6$ Monte Carlo steps. The
simulations were performed near the effective (finite size) critical
temperature found in the previous Section. The histogram method is then used
to extrapolate the needed quantity for different temperatures in the
vicinity of the critical temperature. We follow the same method used in
Refs. \onlinecite{lgk91,lkg91} for the SFXY and TFXY models. The
thermodynamic critical temperature $T_c$ can be determined by the crossing
of the free-energy barriers $\Delta F(T,L)$, obtained from the chirality
histogram $N(\sigma )$ as $\Delta F=A_M(T,L)-A_m(T,L)$, where $A_M$ is the
maximum and $A_m$ one of the two minima in $A(\sigma )=-\log N(\sigma )$. At
the critical point $\Delta F$ is scale invariant but sufficiently close to $%
T_c$, $\Delta F$ can be expanded to linear order in $tL^{1/\nu }$ as $\Delta
F=a+btL^{1/\nu }$, where $t=(T-T_c)/T_c$. In this scaling regime, the
exponent $\nu _I$ can be extracted from the finite-size behavior of the
temperature derivative $\frac{\partial \Delta F}{\partial T}=bL^{1/\nu }$ as
a one-parameter fit in a log-log plot, without requiring a precise
(simultaneous) determination of $T_c$. The exponent $2\beta /\nu $ is
extracted from the scaling behavior of $\sigma _{min}\sim L^{-\beta /\nu }$,
corresponding to the minimum $A_m(\sigma _{\min })$, which only holds at the
critical point and thus is more affected but the estimate of $T_c$.

\begin{figure}[h]
\centering\epsfig{file=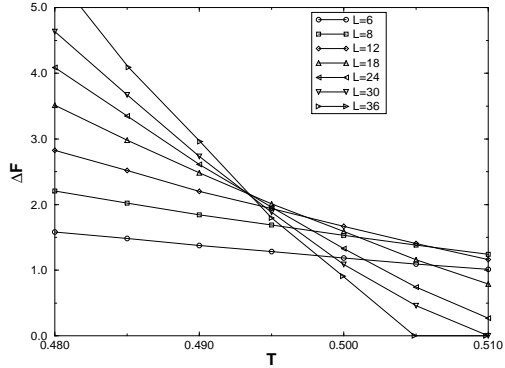,angle=-90,bbllx=1cm,bblly=1cm,bburx=20cm,
bbury=28cm,width=8cm}

\caption{Finite-size scaling of the free energy barrier $\Delta F$ for $\rho
=1.5$. }
\label{cross15.eps}
\end{figure}

\begin{figure}[h]
\centering\epsfig{file=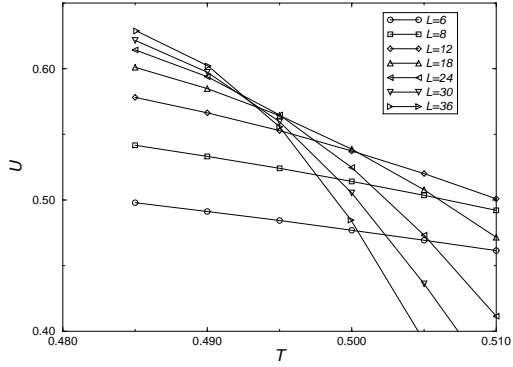,angle=-90,bbllx=1cm,bblly=1cm,bburx=20cm,
bbury=28cm,width=8cm}

\caption{Finite-size scaling of Binder's cumulant $U_L$ for $\rho=1.5$. }
\label{binder.eps}
\end{figure}

\begin{figure}[h]
\centering\epsfig{file=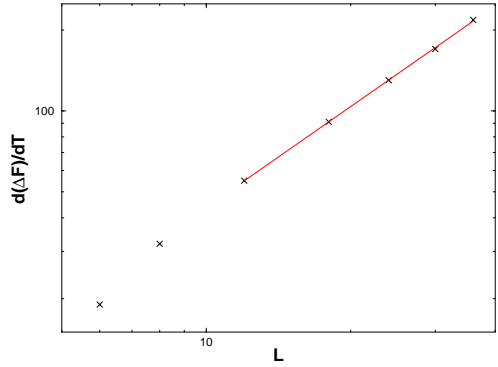,angle=-90,bbllx=1cm,bblly=1cm,bburx=20cm,
bbury=28cm,width=8cm}

\caption{Finite-size scaling of $\frac{\partial \Delta F}{\partial T}$ for $%
\rho =1.5$.}
\label{dfdt15.eps}
\end{figure}

\begin{figure}[h]
\centering\epsfig{file=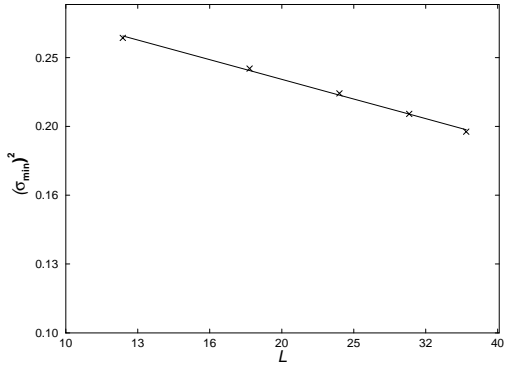,angle=-90,bbllx=1cm,bblly=1cm,bburx=20cm,
bbury=28cm,width=8cm}

\caption{Finite-size scaling of $\sigma_{min}$ for $\rho=1.5$. }
\label{secexp15.eps}
\end{figure}

We have studied two different values of $\rho $ in detail, $\rho =0.7$ and $%
\rho =1.5$, which are located between the SFXY model limit and the Lifshitz
point and between the SFXY model and the TFXY model limit, respectively. For 
$\rho =1.5$ we observed crossing of $\Delta F$ for $L\geq 18$ as shown in
Fig. \ref{cross15.eps}. Corrections to scaling are clearly seen for $6\leq
L\leq 12$. These sizes were not used for the estimates of critical
exponents. Note that, this free energy barriers suffer less from corrections
to scaling than Binder's cumulant \cite{binder}, $U_L=1-<\sigma
_L^4>/3<\sigma _L^2>^2$, which is also expected to cross at a unique point.
This is shown in Fig. \ref{binder.eps} where a sign of unique crossing is
only observed at the largest system sizes. The latter behavior has been used
by Olsson \cite{olsson} in relation to the SFXY model to suggest that there
are in fact two separate transitions and the estimates of $\nu $ are still
dominated by small system sizes. The method we are using, however, indicates
clearly a single crossing point suggesting a reliable estimate of $\nu _I$.
Fig. \ref{dfdt15.eps} shows the size dependence of the slope $\frac{\partial
\Delta F}{\partial T}\vert_{T_c}$ from where $1/\nu =1.25(1)$ can be
estimated and Fig. \ref{secexp15.eps} shows the behavior of $\sigma _{min} $
which gives the estimate $2\beta /\nu =0.29(2)$. The critical temperature is
obtained from the value of $T$ at the crossing point in Fig. \ref{binder.eps}
and gives $T_c = 0.4935(5)$. The same analysis has been done for $\rho =0.7$
giving $T_c=0.408(2)$, $1/\nu =1.28(2)$ and $2\beta /\nu =0.32(4)$. These
estimates deviate significantly from the pure Ising values, $1/\nu =1$ and $%
2 \beta/\nu = 1/4$, and suggest a single transition scenario. Moreover,
these values are consistent with those found for the XY-Ising model along
the critical line \cite{lgk91,ngk95}. This is the same behavior found for
the SFXY and TFXY models using the same methods \cite{lgk91}.


\section{Conclusion}

We have introduced a new generalized version of the square-lattice
frustrated XY model where unequal ferromagnetic and antiferromagnetic
couplings are arranged in a zig-zag pattern. One of the main features of the
model is that the ratio between the couplings $\rho $ can be used to tune
the system through different phase transitions and in one particular limit
it is equivalent to the isotropic triangular-lattice frustrated XY model.
The model can be physically realized as a Josephson-junction array with two
different couplings and in a magnetic field corresponding to a half-flux
quanta per plaquette. We used a mean-field approximation, Ginzburg-Landau
expansion and finite-size scaling of Monte Carlo simulations to study the
phase diagram and critical behavior. Mean-field approximation gives a phase
diagram which qualitatively agrees with the one obtained by Monte Carlo
simulations. Depending on the value of $\rho $, two separate transitions or
a transition line with combined $Z_2$ and $U(1)$ symmetries, takes place.
Based on an effective Hamiltonian, we showed that this transition line is in
the universality class of the XY-Ising model and the phase transitions of
the standard SFXY and TFXY models correspond to two different cuts through
the same transition line. Estimates of the chiral ($Z_2$) critical exponents
from a finite-size analysis of Monte Carlo data were found to be consistent
with previous estimates for the SFXY and TFXY models using the same methods.
They also agree with the corresponding values along the critical line of the
coupled XY-Ising model suggesting a possible physical realization of the
XY-Ising model critical line in a frustrated XY model or Josephson-junction
array with a zig-zag coupling modulation.

\appendix

\section*{}

The mean field equations for the zig-zag model can be derived by an analysis
similar to the one used in Ref. \onlinecite{Gabay89}. The corresponding MF
equations are

\begin{equation}
M_i\equiv \left\langle \vec{S}_i\right\rangle _{MF}=R(\beta H_i)\frac{\vec{H}%
_i}{H_i},  \label{Mean-field}
\end{equation}
where $\vec{H}_i=\sum_jJ_{ij}\vec{M}_j$ is the mean field, $%
R(x)=I_1(x)/I_0(x)$ and $\beta =1/kT$ .

To find the MF phase diagram we expand (\ref{Mean-field}) about the
transition temperature $T_c^{MF}$ using $R(x)=\frac 12x+O(x)$ for $%
x\rightarrow 0$, which reduces to

\begin{equation}
\vec M_i=\frac 12\frac 1{T_c^{MF}}\sum_jJ_{ij}\vec M_j.  \label{mftn0}
\end{equation}
It appears that one needs to make an assumption on the form of the solution $%
M_i$ in order to find $T_c^{MF}$. However, if we note the similarity of Eq. (%
\ref{mftn0}) and the zero temperature limit of Eq. (\ref{Mean-field}), we
can identify the transition temperature as 
\begin{equation}
T_c^{MF}=\frac{H^{GS}}2,  \label{tcmf1}
\end{equation}
provided the local field $H_i=\sum_jJ_{ij}\vec M_j$ is independent of the
position. Although this property is not expected to hold in general, it is
satisfied exactly in the ground state found in Sec. III. We then obtain 
\begin{eqnarray}
\rho &\leq &\frac 13\rightarrow T_c^{MF}=(-\rho +3)/2  \label{TcMFzz} \\
\rho &\geq &\frac 13\rightarrow T_c^{MF}=\sqrt{\left( 1+\rho \right)
^3/4\rho }  \nonumber
\end{eqnarray}
If, in addition, we assume that $H_i$ remains independent of $i$ at any
temperature $0<T<T_c^{MF}$ we obtain 
\begin{equation}
\frac{R(\beta H)}H=\frac 1{H^{GS}}.
\end{equation}
This equation, together with Eq. (\ref{Mean-field}), shows that the
structure of the local configuration around a plaquette and the pitch of the
helical configuration is independent of the temperature in this
approximation.

For $\rho \rightarrow +\infty $ we expect to retrieve the mean field
solution of the TFXY model. However Eq. (\ref{TcMFzz}) leads to a diverging
value of $T_c^{MF}$ as $\rho \rightarrow +\infty $. As can be seen from Eq. (%
\ref{Mean-field}), the temperature is scaled by the magnitude of the mean
field vector $(\left| \sum_jJ_{ij}\vec M_j\right| )$ which diverges when $
\rightarrow +\infty $. This is an artifact of the mean field approximation
and other methods, such as perturbative or variational approximation \cite
{benak2} , can remove this divergence. In fact, the phase diagram obtained
by Monte Carlo simulations in Sec. V leads to a transition temperature that
saturates, for $\rho \rightarrow +\infty $, to a value consistent with the
transition temperature of the TFXY model.

\end{document}